\documentclass[12pt]{article}
\usepackage{amsmath, amsthm, amssymb, amstext}
\usepackage[utf8]{inputenc}
\usepackage[T1]{fontenc}
\usepackage{array,amsfonts,mathrsfs,bm}
\newtheorem{theorem}{Theorem}[section]

\newtheorem{lemma}[theorem]{Lemma}

\title{Stability of determination of Riemann surface from its DN-map in terms of  Teichm\"uller distance}
\author{M.I.Belishev\thanks {St.Petersburg Department of Steklov Mathematical Institute, St.Petersburg, Russia,
        \newline
        e-mail: belishev@pdmi.ras.ru,
        \newline
        ORCID: 0000-0002-4759-7428;
        \newline
        supported by RFBR grant 20-01-00627-a
        \newline }.\,
        D.V.Korikov\thanks {St.Petersburg Department of Steklov Mathematical Institute, St. Petersburg, Russia,
        \newline
        e-mail: thecakeisalie@list.ru,
        \newline
        ORCID: 0000-0002-3212-5874;
        \newline
        supported by RFBR grant 20-01-00627-a
        \newline }.}

\date{}

\begin{document}
\maketitle

\begin{abstract}
As is known, the Dirichlet-to-Neumann operator $\Lambda$ of a Riemannian surface $(M,g)$ determines the surface up to conformal equivalence class $[(M,g)]$. Such classes constitute the Teichm\"uller space with the distance ${\rm d}_T$. We show that the determination is continuous: $\|\Lambda-\Lambda'\|_{H^1(\partial M)\to L_2(\partial M)}\to 0$
implies ${\rm d}_T([(M,g)],[(M',g')])\to 0$.

\end{abstract}

\noindent{\bf Key words:}\,\,\,electric impedance tomography of
surfaces, holomorphic immersions, Di\-rich\-let-to-Neumann map,
stability of determination.


\bigskip

\section*{Introduction}
\subsubsection*{Two-dimensional EIT problem}
$\bullet \ $ Let $(M,g)$ be a smooth\footnote{throughout the paper, {\it smooth} means $C^\infty$-smooth} compact two-dimensional Riemann manifold (a {\it surface}) with the smooth boundary $(\Gamma,dl)$, $g$ the smooth metric tensor. For the sake of simplicity, we assume that $\Gamma$ is diffeomorphic to a circle,
and denote by $dl$ and $\nu$ the length element and unit normal on $\Gamma$. Let $\Delta_g$ be the Beltrami-Laplace operator on $M$.

The {\it Dirichlet-to-Neumann operator} (DN map) of the surface $(M,g)$ acts on smooth functions on $\Gamma$ by the rule $\Lambda
f:=\partial_{\nu}u^{f}\big|_{\Gamma}$, where $u^{f}$ satisfies
$$ \Delta_g u=0\quad\text{in}\,\,M\setminus\Gamma;\qquad u=f\quad\text{on}\,\,\Gamma.$$
The {\it electric impedance tomography problem}(EIT) is to determine the surface from its DN map. Here are some known facts and results.
\smallskip

$\bullet \ $ 
Suppose that $(M,g)$ and $(M',g')$ are the surfaces with common boun\-dary\footnote{in the subsequent, the term `common boundary' means that the surface metrics induce the same length element on the boundary.} $\partial M=\partial M'=\Gamma$; let $\Lambda$ and $\Lambda'$ be their DN maps. We write $(M',g')\sim (M,g)$ if they are conformally equivalent via a diffeomorphism 
that does not move points of the boundary. By $[(M,g)]$ we denote the equivalence class of $(M,g)$. 

In \cite{LU} M.Lassas and G.Uhlmann showed that $\Lambda$ uniquely determines not the metric $g$ on $M$ but its conformal class, so that the correspondence $\Lambda\leftrightarrow[(M,g)]$ is a bijection. Also, a procedure that recovers $M$ by means of harmonic continuation from the boundary, is provided.
In \cite{B}, the same fact is established by algebraic
version of the boundary control method (BCM) \cite{B UMN}. Recently, it was extended to the case of non-orientable surfaces \cite{BKor_JIIPP,BKor_SIAM} and surfaces with (unknown) internal holes \cite{BKor_IP}. 

The paper \cite{H&M} by G.Henkin and V.Michel provides a constructive way to recover the surface from its DN map by the use of multidimensional complex analysis technique. Moreover, a characteristic description of the inverse data is given that is the necessary and sufficient conditions on an operator $\Lambda$ to be the DN map of a surface. In \cite{BKor_CAOT} another simpler characterization
in the framework of the algebraic BCM is provided. 
\smallskip

$\bullet \ $ The question on {\it stability of determination} $\Lambda\mapsto(M,g)$ can be posed as follows. Let the operators $\Lambda$ and $\Lambda'$ that correspond to the surfaces $(M,g)$ and $(M',g')$, be close (in a relevant sense). Can one claim that the surfaces are also close (in appropriate sense)? An affirmative answer given in our paper \cite{BKstab} is as follows.

First, as is clarified in \cite{ZNS}, to discuss a closeness of the surfaces is reasonable only under assumption that $M$ and $M'$ are diffeomorphic, i.e., have the same genus $m\geqslant 0$. Accepting this for the rest of the given paper, we deal with the {\it diffeomorphic} surfaces $(M,g)$ and $(M',g')$ with the common boundary $\Gamma$ and the length element $dl$ on it. Also, the surfaces are assumed orientable and oriented in accordance with a fixed orientation of $\Gamma$.

Let $\mathcal{E}: M\mapsto\mathbb C^n$ be a 
holomorphic immersion of $(M,g)$; let us write $M'\in\mathbb M_t$ if
$\parallel\Lambda'-\Lambda\parallel_{H^{1}(\partial M;\mathbb R)\to L_{2}(\partial M;\mathbb R)}\leqslant t$ is 
fulfilled. Then the convergence
$$\sup_{M'\in \mathbb{M}_{t}}\inf_{\mathcal{E}'}d_{H}(\mathcal{E}'(M'),\mathcal{E}(M))\underset{t\to 0}{\longrightarrow}0$$ 
holds, 
where $d_H$ is the Haussdorf distance in $\mathbb C^n$ and the 
infimum is taken over all holomorphic immersions $\mathcal{E}': M'\mapsto\mathbb C^n$ \cite{BKstab}.

\subsubsection*{Teichm\"uller metric} 
$\bullet \ $ The motivation to make the latter result more expressive and natural is as follows. As was noted above, the DN map $\Lambda$ determines not the metric $g$ on $M$ but the conformal equivalence class of metrics $[(M,g)]$. Such classes constitute the Teichm\"uller space $\mathfrak T_m$ \cite{T} (the subscript $m$ indicates the genus of the surfaces). This space is endowed with the natural Teichm\"uller metric defined below. The set $\mathscr{D}_m$ of the DN maps of surfaces of genus $m$ is contained in the normed space $\mathfrak B_{10}$ of the bounded linear operators, which act from
$H^1(\Gamma;\mathbb R)$ to $H^0(\Gamma;\mathbb R):=L_2(\Gamma;\mathbb R)$. Therefore, it would be most relevant to consider a continuity of the correspondence $\Lambda\mapsto [(M,g)]$ from $\mathscr{D}_m$ to $\mathfrak T_m$. Such a continuity is the main result of our work. To present it, we begin with basic notions and known facts.
\smallskip

$\bullet$\ Let $(M,g)$ and $(M',g')$ be the surfaces and  let $q: \ M\to M'$ be an orientation preserving diffeomorphism. In the holomorphic local coordinates $z$ on $M$ and $z'$ on $M'$, its differential is of the form
$$dq(z)=\partial_z q(z)\left[\,dz+\mu(z)d\overline{z}\,\right],$$
where 
$$\mu(z):=\frac{\partial_{\overline{z}}q(z)}{\partial_z q(z)}$$ 
is called the {\it Beltrami quotient} of $q$. The Jacobian of $q$ obeys ${\rm Jac\,}(q)=|\partial_z q|^2-|\partial_{\overline{z}}q|^2>0$, whence $|\mu(z)|<1$. Also, $|\mu(z)|$ does not depend on the choice of holomorphic coordinates $z$ an $z'$. Thus, the function
$$k(x):=\frac{1+|\mu(z)|}{1-|\mu(z)|}$$
(where $z$ is a holomorphic coordinate of $x$) is continuous on $M$. Its value $k(x)$ is called a {\it dilatation} of the map $q$ at the point $x$. Note that $k(x)$ is equal to the square of the ratio between length of major and minor axis of the ellipse by pulling back along $q$ the unit circle in $T_{q(x)}M'$. The map $q$ is called $K$-{\it quasi-conformal} if its dilatation obeys $\sup_{x\in M}k(x)=K$; in this case the number $K$ is called the dilatation of $q$.

If $(M,g)$ and $(M',g')$ are conformally equivalent (belong to the same class $[(M,g)]$) via a diffeomorphism $q$
then one has $\partial_{\bar z}q=0,\,\,\,\mu(\cdot)=0,\,\,\,k(\cdot)=1$, and $K=1$. Otherwise, the value $K\not=1$ shows to what extent the map $q$ differs from being conformal. This motivates the basic definition of the Teichm\"uller distance: for $[(M,g)],[(M',g')]\in\mathfrak T_m$ one puts
 \begin{equation}
\label{TM}
{\rm d}_T([(M,g)],[(M',g')]):=\frac{1}{2}\,{\rm log}\inf_{q}K(q),
\end{equation}
where the infimum is taken over all orientation preserving diffeomorphisms $q: \ M\to M'$ and $K(q)$ denotes the dilatation of $q$. Note that $d_T([(M,g)],[(M',g')])$ does not depend on the choice of elements of conformal classes.
\smallskip

%$\bullet$\ 
%In spirit, ${\rm d}_T$ is close to the Banach-Masur distance between finite dimensional normed spaces. Namely, $${\rm d}_{BM}{(\mathscr L,\mathscr M)}:=\log\inf\left\{\,\|T^{-1}\|\|T\|\,\,|\,\,T\,\,\text{is an isomorphism from}\,\,\mathscr L\,\,\text{onto}\,\,\mathscr M\right\},$$ where the product of the norm shows, to what extent $T$ differs from being an isometry between the spaces.

\subsubsection*{Main result} 
$\bullet \ $ Recall that $\mathfrak T_m$ is the set of conformal equivalence classes $[(M,g)]$ of the genus $m$ oriented surfaces $(M,g)$ with the common boundary $\Gamma$ and length element $dl$ on it. The surfaces $(M,g)\in[(M,g)]$ 
have the same DN map $\Lambda$. 
For a fixed class $[(M,g)]$, by $\mathfrak M_t\subset\mathfrak T_m$ we denote the set of the classes $[(M',g')]$ provided $\|\Lambda-\Lambda'\|_{\mathfrak B_{10}}\leqslant t$, where $\Lambda'$ is DN map of $(M',g')$. The result of the paper is the following.
\begin{theorem}
\label{MT}
Let $[(M,g)]\in \mathfrak{T}_{m}$ and $\Lambda$ be the DN map of $[(M,g)]$. Then
the relation
\begin{equation}
\label{MTf}
\sup_{[(M',g')]\in\mathfrak{M}_t}{\rm d}_T([(M,g)],[(M',g')])\stackrel{t\to 0}{\longrightarrow}0.
\end{equation}
holds and means that the correspondence $\Lambda\mapsto[(M,g)]$ is continuous from $\mathscr{D}_m$ to $\mathfrak T_m$.
\end{theorem}
The remainder of the paper is devoted to the proof of Theorem \ref{MT}. Its idea is to provide constructively a map $\alpha: \ (M,g)\to (M',g')$, whose dilatation is close to 1 uniformly with respect to $(M',g')$, the closeness being derived from the closeness of $\Lambda$ and $\Lambda'$.

\section{Preliminaries}
In this section, we recall the basic notions and facts along with the results of \cite{BKstab} that will be used in the proof of Theorem \ref{MT}. In the rest of paper, the objects associated with the surface $M$ are designated by unprimed symbols, while objects associated with the surface $M'$ are designated by primed symbols.
\subsubsection*{Holomorphic functions} 
$\bullet \ $ Since the the surface $(M,g)$ is orientable, there is the smooth family $\Phi=\{\Phi_x\}_{x\in M}$ of `rotations' $\Phi_x\in {\rm End}T_x M$ such that
$$g(\Phi_x a,\Phi_x b)=g(a,b), \quad g(\Phi_x a,a)=0, \qquad \forall a,b\in T_x M, \ x\in M.$$
The boundary $\Gamma$ of $M$ is oriented by the tangent field $\gamma:=\Phi\nu$. In the subsequent, dealing with the set of surfaces $(M,g)$, $(M',g')$, $\dots$ with the common boundary $(\Gamma,dl)$, we agree that their orientations are consistent in such a way that $\Phi\nu=\Phi'\nu'=\dots=\gamma$.

A smooth function $w: \ M\mapsto\mathbb{C}$ is holomorphic if the Cauchy-Riemann condition $\nabla_g\Im w=\Phi\nabla_g\Re w$ holds in $M$; then its real and imaginary parts are harmonic i.e. $\Delta_g \Re w=\Delta_g \Im w=0$ holds in $M$. Denote the lineal of holomorphic smooth functions on $M$ by $\mathscr{H}(M)$. 

$\bullet \ $ The one instrument in the proof of Theorem \ref{MT} is the {\it generalized argument principle}: 
\begin{equation}
\label{GAP}
\frac{1}{2\pi i}\int\limits_{\Gamma}\tilde{w}\frac{dw}{w-z}=\sum_{x\in w^{-1}(\{z\})}{\rm ord}_x(w-z)\tilde{w}(x)
\end{equation}
for any $w,\tilde{w}\in\mathscr{H}(M)$ and $z\in\mathbb{C}\backslash\eta(\Gamma)$; here ${\rm ord}_x(w-z)$ is the order of zero $x$ of the function $w-z$. Formula (\ref{GAP}) follows from Stokes theorem (see Theorem 3.16, \cite{M}) and the residue theorem (see Lemma 3.12, \cite{M}) for meromorphic
1-form $\tilde{w}dw/(w-z)$. If $\tilde{w}=1$, then the right-hand side of (\ref{GAP}) is just a total multiplicity ${\rm mul}(w-z)$ of zeroes of $w-z$ and (\ref{GAP}) becomes a usual argument principle. 

$\bullet \ $ The trace operator ${\rm Tr}: \ w\mapsto w|_{\Gamma}$ is a bijection from $\mathscr{H}(M)$ to its image ${\rm Tr}\mathscr{H}(M)$. Let us characterize the lineal ${\rm Tr}\mathscr{H}(M)$ in terms of DN-map $\Lambda$. Denote the differentiation on the boundary $\Gamma$ with respect to the tangent field $\gamma$ by $\partial_\gamma$. Also, denote the orthogonal complement to constants in $L_2(\Gamma;\mathbb{R})$ by $\dot{L}_2(\Gamma;\mathbb{R})$ and introduce the integration $J=\partial_{\gamma}^{-1}$ on $\dot{L}_2(\Gamma;\mathbb{R})$. 

As is known, the DN map $\Lambda$ is a first-order pseudo-differential operator that acts continuously from $H^m(\Gamma,\mathbb R)$ to $H^{m-1}(\Gamma,\mathbb R),\,\,\,m=1,2,\dots$, where $H^m(\dots)$ are Sobolev spaces. Its closure in $L_2(\Gamma;\mathbb R)$ (still denoted by $\Lambda$) obeys ${\rm Ker\,}\Lambda=\{\rm const\}$ and $\overline{{\rm Ran\,}\Lambda}=\dot{L}_2(\Gamma,\mathbb R)$. In particular, $J\Lambda$ is well defined bounded operator in each $H^m(\Gamma,\mathbb R),\,\,\,m=1,2,\dots$. Similarly, $\Lambda J$ is well defined bounded operator in each $H^m(\Gamma,\mathbb R)\cap \dot{L}_2(\Gamma,\mathbb R),\,\,\,m=0,1,\dots$. 

Denote the projector in $L_2(\Gamma,\mathbb R)$ onto the subspace ${\rm Ker}[\mathbb{I}+(\Lambda J)^2]^{*}\oplus\mathbb{R}$ by $P$ (here $\mathbb{I}$ is the unit operator in $L_2(\Gamma;\mathbb{R})$). As is shown in \cite{B}, the formula
\begin{equation}
\label{traces}
{\rm Tr}\mathscr{H}(M)=\{ Pf+i[J\Lambda Pf+c] \ | \ f\in C^{\infty}(\Gamma;\mathbb{R}), \ c\in\mathbb{R}\}
\end{equation}
is valid. Also, the dimension of $(I-P)L_2(\Gamma,\mathbb R)$ is finite and it is equal to $2 \ {\rm gen}(M)$; hence, the projector $P$ determines the topology of $M$. In particular, if $M$ is a disc in $\mathbb{R}^2$, then $P=\mathbb{I}$ and $J\Lambda$ coincides with the classical Hilbert transform that relates real and imaginary parts of traces of holomorphic functions. 

$\bullet \ $ Let $(M',g')\in\mathbb{M}_t$ be a surface with DN-map $\Lambda'$. Due to (\ref{traces}), we can define a real linear map $\hat{\beta}_{M'}: \ {\rm Tr}\mathscr{H}(M)\to{\rm Tr}'\mathscr{H}(M')$ by the rule
$$\hat{\beta}_{M'}\eta=P'\Re\eta+i[J\Lambda P'\Re\eta+\langle\Im\eta\rangle],$$
where $\langle f\rangle:=\int_\Gamma fdl$, and $P'$ is a projector in $L_2(\Gamma,\mathbb R)$ onto the subspace ${\rm Ker}[\mathbb{I}+(\Lambda' J)^2]^{*}\oplus\mathbb{R}$. In \cite{BK}, it is proved that, for sufficiently small $t\in [0,t_0)$ and any $M'\in\mathbb{M}_t$, the map $\hat{\beta}_{M'}$ is bijective and the estimate
\begin{equation}
\label{beta est}
\|\hat{\beta}_{M'}\eta-\eta\|_{C^m(\Gamma;\mathbb{C})}\le c_m t\|\eta\|_{H^{m+1}(\Gamma;\mathbb{C})}, \qquad \forall\eta\in{\rm Tr}\mathscr{H}(M)
\end{equation}
is valid for $m=1,2,\dots$. Here and in the subsequent, all constants in the estimates are assumed to be independent of $t$ and $M'\in\mathbb{M}_t$. We introduce the `canonical' real linear map 
\begin{equation}
\label{beta map}
\begin{split}
\beta_{M'}: \ \mathscr{H}(M)\to\mathscr{H}(M'), \qquad \beta_{M'}={\rm Tr}^{'-1}\circ\hat{\beta}_{M'}\circ{\rm Tr}.
\end{split}
\end{equation}

\subsubsection*{Holomorphic embeddings} 
$\bullet \ $ Recall that an immersion is a differentiable map $\alpha : \ M_1\mapsto M_2$ between two differentiable manifolds $M_1$ and $M_2$, whose differential $d\alpha_{x}: \ T_x M_1\mapsto T_{\alpha(x)} M_2$ is injective for any $x\in M_1$. The immersion $\alpha$ is an {\it embedding} if $\alpha: \ M_1\mapsto \alpha(M_1)$ is a homeomorphism, (where the topology on $\alpha(M_1)$ is induced by $M_2$). We say that the embedding
$$\mathcal{E}: \ M\to\mathbb{C}^n, \quad x\mapsto \{w_1(x),\dots,w_n(x)\} $$
is {\it holomorphic} if it is realised by holomorphic functions $w_k$. In what follows, we denote $\eta_k:={\rm Tr}w_{k}$.

For $\xi=\{\xi_1,\dots,\xi_n\}\in \mathbb{C}^n$, introduce the coordinate projections $\pi_i: \ \xi\mapsto \xi_i$. Let $D$ be a domain in $\mathbb{C}$, denote the cylinder $\{\zeta\in\mathbb{C}^n \ | \ \pi_i\zeta\in D\}$ by $\Pi_{i}[D]$. We say that $\Pi_{i}[D]$ is $\mathcal{E}(M)$-projective if $\pi_i: \ \Pi_{i}[D]\cap\mathcal{E}(M)\to\mathbb{C}$ is an embedding. The embedding $\mathcal{E}$ is called {\it projective} if each point $\xi\in\mathcal{E}(M)$ belongs to some $\mathcal{E}(M)$-projective cylinder. The existence of projective embeddings follows from the divisor theorem (see Theorem ?, \cite{Forster}). In the parper, we deal with projective holomorphic embeddings only and, for short, call them just `embeddings'. The basic properties of such embeddings are presented below.

$\bullet \ $ The image of an embedding of a surface is determined only its DN-map and the choice of boundary traces $\eta_k$. Indeed, let $(M,g)$ and $(M',g')$ be surfaces with the joint boundary $(\Gamma,dl)$ and their DN-maps $\Lambda$ and $\Lambda'$ coincide. Then there exists a conformal map $\beta: \ M\to M'$ that does not move the points of $\Gamma$. Then the functions $w_k\in\mathscr{H}(M)$, $w'_k\in\mathscr{H}(M')$ with joint boundary traces $\eta_k:={\rm Tr}w_k={\rm Tr}'w'_k$ are connected by $w_{k}=w'_{k}\circ\beta$. Hence, $\mathcal{E}(M)=\mathcal{E}'(M)$ is valid for $\mathcal{E}=\{w_1,\dots,w_n\}$ and $\mathcal{E}'=\{w'_1,\dots,w'_n\}$.

$\bullet \ $ Let $\mathcal{E}=\{w_1,\dots,w_n\}$ be an embedding of $(M,g)$. The surface $\mathcal{E}(M)$ is endowed with the metric $\tilde{g}$ induced by standard metric in $\mathbb{C}^n$. With this choice of metric, the map $\mathcal{E}: \ M\to \mathcal{E}(M)$ is conformal. Indeed, suppose that $\phi: \ [0,1]\to M$ is a smooth curve with the beginning at $x_0$ and the tangent vector $\theta$ at $x_0$. Let $\tilde{\phi}=\mathcal{E}\circ\theta$ be a corresponding curve in $\mathcal{E}(M)$ with the beginning at $\xi_0=\mathcal{E}(x_0)$ and the tangent vector $\tilde{\theta}=d\mathcal{E}_x \theta$ at $\xi$. Chose a projective cylinder $\Pi_i[D]$ containing $\xi$, then $(\mathcal{E}(M)\cap\Pi_i[D],\pi_i)$ is a complex chart on $\mathcal{E}(M)$ and $(\mathcal{E}^{-1}(\Pi_i[D]),w_i=\mathcal{E}\circ\pi_i)$ is a complex chart on $M$. In local coordinates $(x^1,x^2)=(\Re w_i(x),\Im w_i(x))=(\Re\pi_i\xi,\Im\pi_i\xi)$, where $\xi=\mathcal{E}(x)$, the components of vectors $\theta$ and $\tilde{\theta}$ satisfy
$$\theta^1+i\theta^2=\frac{dw_i\circ\phi}{dt}(0)=\frac{d\pi_i\circ\mathcal{E}\circ\phi}{dt}(0)=\frac{d\pi_i\circ\tilde{\phi}}{dt}(0)=\tilde{\theta}^1+i\tilde{\theta}^2.$$
Since $w_i$ is holomorphic, we have $g_{kl}=\rho_1(x)\delta_{kl}$ in local coordinates $(x^1,x^2)$, where $\rho_1$ is a conformal factor. Hence, $g(\theta,\theta)=\rho_1(x_0)|\theta^1+i\theta^2|^2$. Since the metric $\tilde{g}$ on $\mathcal{E}(M)$ is induced by $\mathbb{C}^n$, we have 
$$\tilde{g}(\tilde{\theta},\tilde{\theta})=\sum_{k=1}^n \Big|\frac{dw_k\circ w_{i}^{-1}}{dz}\frac{d\pi_i\circ\tilde{\phi}}{dt}(0)\Big|^2=\rho_2(x_0)|\tilde{\theta}^1+i\tilde{\theta}^2|^2,$$
where $z=\pi_i\xi=w_i(x)$ and $\rho_2(x)=\sum_{k=1}^n |\partial_z w_k\circ w_{i}^{-1}|^2$. Hence, we have  $\tilde{g}(\tilde{\theta},\tilde{\theta})=\rho_2(x_0)\rho_1(x_0)^{-1}g(\theta,\theta)$ and, since $\phi$ is arbitrary, $\tilde{g}=\rho\mathcal{E}^{*}g$, where $\rho\in C^{\infty}(\mathcal{E}(M);(0,+\infty))$.

Note that the collection $\{U=\mathcal{E}^{-1}(\Pi_i[D]),w_i|_U\}$ (where $\Pi_i[D])$ is $\mathcal{E}(M)$-projective cylinder) provides a byholomorphic atlas on $(M,g)$ whereas the collection $\{V=\mathcal{E}(M)\cap\Pi_i[D]),\pi_i|_V\}$ provide a byholomorphic atlas on $(\mathcal{E}(M),\tilde{g})$ (both atlases are consistent with the metrics). With such atlases, the map $\mathcal{E}: \ M\to\mathcal{E}(M)$ is byholomorphism.

$\bullet \ $ For the embedding $\mathcal{E}=\{w_1,\dots,w_n\}$, application of the generalized argument principle (\ref{GAP}) yields the following. Let $\xi=\mathcal{E}(x)$ belongs to some projective cylinder $\Pi_i[D]$ and $z=\pi_i x$. Since the map $\xi\to\pi_i(x)$ is an embedding, $x$ is a unique and simple zero of the function $w_i-z=\pi_i\circ\mathcal{E}-z$. Due to (\ref{GAP}), the projection $\pi_k \xi=w_k(x)$ can be found by
\begin{equation}
\label{GAPemb}
\pi_k \xi=\frac{1}{2\pi i}\int\limits_{\Gamma}\eta_k\frac{\partial_{\gamma}\eta_i}{\eta_i-z}dl.
\end{equation}
So, the conformal copy $\mathcal{E}(M)$ of $M$ can be reconstructed from $\{\eta_{1},\dots,\eta_n\}$ by using (\ref{GAPemb}).

\subsubsection*{Induced embeddings}

$\bullet \ $ Let $\mathcal{E}=\{w_1,\dots,w_n\}$ be a (fixed) embedding of $(M,g)$. For a surface $(M',g')\in\mathbb{M}_t$, we say that its embedding $\mathcal{E}'=\{w'_1,\dots,w'_n\}$ is {\it induced by} $\mathcal{E}$ if the functions $w'_k$ are connected with $w_k$ by $w'_k=\beta_{M'}w_k$, where the map $\beta_{M'}$ is defined in (\ref{beta map}). Then (\ref{beta est}) yields
\begin{equation}
\label{b tr est}
\|\eta'_k-\eta_k\|_{C^{m}(\Gamma;\mathbb{C})}\le c_m t \qquad (m=1,2,\dots)
\end{equation}
for boundary traces $\eta_k={\rm Tr}w_k$, $\eta'_k={\rm Tr}'w'_{k}$. 

In the subsequent, $\mathcal{E}'$ always denotes the embedding of $(M',g')$ induced by $\mathcal{E}$. In \cite{BK}, it is proved that, for sufficiently small $t\in [0,t_0)$ and any $(M',g')\in\mathbb{M}_t$, the induced embedding $\mathcal{E}'$ is actually projective and 
$$\sup_{(M',g')\in\mathbb{M}_t}d_H(\mathcal{E}'(M'),\mathcal{E}(M))\underset{t\to 0}{\to} 0,$$
where $d_H(K_1,K_2)$ is the {\it Hausdorff distance} between two compacts $K_1$ and $K_2$ in $\mathbb{C}^n$ defined as the infimum of positive $s$ such that $s$-neighbourhood of $K_1$ contains $K_2$ and $s$-neighbourhood of $K_2$ contains $K_1$.

\section{The map $\alpha$}

$\bullet$ \ Assume that $(M,g)$ is a surface with boundary $(\Gamma,dl)$ and $\mathcal{E}$ is a fixed embedding of $(M,g)$. Also, suppose that $(M',g')\in\mathbb{M}_t$ with small $t>0$ and $\mathcal{E}'$ is an embedding of $(M',g')$ induced by $\mathcal{E}$. In this section, we construct the near-isometric diffeomorphism $\alpha=\alpha_{M'}$ from $\mathcal{E}(M)$ onto $\mathcal{E'}(M')$. Informally speaking, the map $\alpha$ is defined is the following way. If $\xi\in\mathcal{E}(M)$ is separated from $\mathcal{E}(\Gamma)$, then $\alpha(\xi)$ is a point $\xi'$ of $\mathcal{E}'(M')$ closest to $\xi$ in $\mathbb{C}^n$. If $\xi\in \mathcal{E}(M)$ is close to $\mathcal{E}(\Gamma)$, then $\alpha(\xi)$ is the point $\xi'\in\mathcal{E}'(M')$ whose semi-geodesic coordinates $(l',r')$ on $\mathcal{E}'(M')$ coincides with the semi-geodesic coordinates $(l,r)$ of $\xi$. 

$\bullet$ \ More precisely, $\alpha(\xi)$ is defined as the point of minimum of a certain function $\mathfrak{D}(\xi,\cdot): \ \mathcal{E}'(M')\mapsto[0,+\infty)$ which is constructed in the following way. First, introduce the function
\begin{equation}
\label{d int}
\mathfrak{D}_{int}: \ \mathcal{E}(M)\times\mathcal{E}'(M')\mapsto [0,+\infty), \qquad \mathfrak{D}_{int}(\xi,\xi'):=|\xi'-\xi|^2;
\end{equation}
then $\mathfrak{D}_{int}(\xi,\cdot)$ attain the minimum at a point $\xi'$ of $\mathcal{E}'(M')$ closest to $\xi$. 

For $\xi\in\mathcal{E}(M)$, let $r$ stands for distance between $\xi$ and $\mathcal{E}(\Gamma)$ in $(\mathcal{E}(M),\tilde{g})$. Also, let $l:=\mathcal{E}^{-1}(\zeta)$, where $\zeta$ is a point of $\mathcal{E}(\Gamma)$ closest to $\xi$ in $(\mathcal{E}(M),\tilde{g})$. The pair $(l,r)$ provides semi-geodesics coordinates on $(\mathcal{E}(M),\tilde{g})$, which are regular at least on a certain near-boundary strip $r\le c$. The semi-geodesics coordinates $(l',r')$ on $(\mathcal{E}'(M'),\tilde{g}')$ are defined in the same way. In what follows, we show that, for sufficiently small $t$, the coordinates $(l',r')$ are regular on the near-boundary strip $r'\le r_0$, where $r_0>0$ does not depend on $t$ and $M'$. Introduce the function
\begin{equation}
\label{d gamma}
\begin{split}
\mathfrak{D}_{\Gamma}&: \ \mathcal{E}(M)\times\mathcal{E}'(M')\mapsto [0,+\infty), \\
\mathfrak{D}_\Gamma &(\xi,\xi'):=({\rm dist}_{\Gamma}(l'(\xi'),l(\xi))^2+(r'(\xi')-r(\xi))^2;
\end{split}
\end{equation}
then $\mathfrak{D}_{\Gamma}(\xi,\cdot)$ ($r(\xi)<r_0$) attain the minimum at a point $\xi'$ with semi-geodesics coordinates $l'=l$, $r'=r$.

Let $\kappa$ be a non-negative smooth cut-off function on $\mathcal{E}(M)$ which is equal to zero for $r>2r_0 /3$ and to one for $r<r_0 /3$. The function $\mathfrak{D}: \ \mathcal{E}(M)\times\mathcal{E}'(M')\to [0,+\infty)$ is defined by
\begin{equation}
\label{d whole}
\mathfrak{D}:=(1-\kappa)\mathfrak{D}_{int}+\kappa \mathfrak{D}_{\Gamma}.
\end{equation}

$\bullet$ \ Along with $\mathfrak{D}$, we also consider the `unperturbed' function
\begin{equation}
\label{d unpert}
\mathfrak{D}_0:=(1-\kappa)\mathfrak{D}_{int,0}+\kappa \mathfrak{D}_{\Gamma,0}
\end{equation}
which coincides with $\mathfrak{D}$ in the case $\mathcal{E}'(M')=\mathcal{E}(M)$. Here the the functions $\mathfrak{D}_{int,0}$, $\mathfrak{D}_{\Gamma,0}$ are defined by (\ref{d int}) and (\ref{d gamma}), where $\mathcal{E}'(M')$ is replaced by $\mathcal{E}(M)$ and $l',r'$ are replaced by $l,r$. Note that $\xi'=\xi$ is a unique and non-degenerated point of global minimum of the function $\xi'\mapsto\mathfrak{D}_0(\xi,\xi')$ ($\xi\in\mathcal{E}(M)$).

\subsection{Construction of $\alpha$ in a zone separated from $\mathcal{E}(\Gamma)$.}
\label{ss out of b}
$\bullet$ \ Let $\Pi_i[D]$ be an $\mathcal{E}(M)$-projective cylinder whose closure does not intersects with $\mathcal{E}(\Gamma)$. Then $|\eta_i(l)-z|>c>0$ and ${\rm mul}(w_i-z)=1$ for $z\in\overline{D}$, $l\in\Gamma$ and (\ref{GAP}), (\ref{b tr est}) yields
$$|{\rm mul}(w'-z)-{\rm mul}(w-z)|=\Big|\frac{1}{2\pi i}\int\limits_{\Gamma}\frac{\partial_{\gamma}\eta'_i}{\eta'_i-z}dl-\frac{1}{2\pi i}\int\limits_{\Gamma}\frac{\partial_{\gamma}\eta'_i}{\eta'_i-z}dl\Big|\le ct.$$
Thus, for sufficiently small $t>0$, we have ${\rm mul}(w'_i-z)=1$, whence the cylinder $\Pi_i[D]$ is also $\mathcal{E}'(M')$-projective. Similarly, differentiating (\ref{GAPemb}) and applying estimate (\ref{b tr est}), we obtain
\begin{equation}
\label{GAP est}
\|w'_j\circ w_{i}^{'-1}-w_j\circ w_{i}^{-1}\|_{C^{m}(D;\mathbb{C})}\le c_m t, \qquad (m=1,2,\dots).
\end{equation}
Assuming that $\kappa=0$ on $\mathcal{E}(M)\cap\Pi_i[D]$ and considering $\mathfrak{D}=\mathfrak{D}_{int}$ and $\mathfrak{D}_0=\mathfrak{D}_{int,0}$ as functions of local coordinates $x^1+ix^2=z=\pi_i \xi$, $x^{'1}+ix^{'2}=z'=\pi_i \xi'$, we obtain
\begin{align*}
\mathfrak{D}(\xi,\xi')&=\mathfrak{D}_{int}(\xi,\xi')=\sum_{k=1}^{n}|w'_j\circ w_i^{'-1}(z')-w_j\circ w_i^{-1}(z)|^2,\\
\mathfrak{D}_0(\xi,\xi')&=\mathfrak{D}_{int,0}(\xi,\xi')=\sum_{k=1}^{n}|w_j\circ w_i^{-1}(z')-w_j\circ w_i^{-1}(z)|^2,
\end{align*}
where $z,z'\in D$. Then (\ref{GAP est}) implies
\begin{equation}
\label{func est}
\|\mathfrak{D}-\mathfrak{D}_0\|_{C^m(D\times D;[0,+\infty))}\le c_m t, \qquad m=1,2,\dots.
\end{equation}
Decreasing the diameter of $D$, we can assume that $\xi'=\xi$ is a unique point of local minimum of the function $\xi'\mapsto\mathfrak{D}_0(\xi,\xi')$ in $\mathcal{E}(M)\cap\Pi_i[D]$. 

$\bullet$ \ Let $\Pi_i[\mathcal{D}]$ is a sub-cylinder of $\Pi_i[D]$ such that $\overline{\mathcal{D}}\subset D$. To construct the map $\xi\mapsto\alpha(\xi)$ for $\xi\in\mathcal{E}(M)\cap \Pi_i[\mathcal{D}]$, we need the following
\begin{lemma}
\label{IFLemma}
Let $X,\mathcal{X},Y$ be domains with compact closures in $\mathbb{R}^{n}$, $\overline{\mathcal{X}}\subset X$, and $F\in C^2(\overline{X}\times \overline{Y};\mathbb{R}^n)$ satisfies {\rm i)} the zero set of $F$ is the graph of the function $f\in C^1(X;Y)\cap C(\overline{X};\overline{Y})$ and {\rm ii)} for any $x\in X$ there exists $(F'_y(x,f(x)))^{-1}$. Then there exists $t_*>0$ such that, for any $t\in (0,t_*)$ and any $H\in C^1(X\times Y;\mathbb{R}^n)$ satisfying
\begin{equation}
\label{FG C1-closeness}
\|H-F\|_{C^1(X\times Y;\mathbb{R}^n)}\le t,
\end{equation}
the zero set of $H$ in $\mathcal{X}\times Y$ is the graph of the function $h\in C^1(\mathcal{X};Y)$ and 
$$\|h-f\|_{C^1(\mathcal{X};Y)}\le ct$$ 
{\rm (}the constant $c$ does not depend on $H${\rm )}. Also, if $f$ is a diffeomorphism from $X$ to $f(X)$, then $h$ is a a diffeomorphism from $\mathcal{X}$ to $g(\mathcal{X})$.
\end{lemma}
Lemma \ref{IFLemma} is just a variant of the implicit functions theorem; for the convenience of the reader, its proof is contained in Appendix \ref{App1}.

In Lemma \ref{IFLemma}, put $X=Y=D$, $\mathcal{X}=\mathcal{D}$, $F=(\partial_{x^{'1}}\mathfrak{D}_0,\partial_{x^{'2}}\mathfrak{D}_0)$, and $H=(\partial_{x^{'1}}\mathfrak{D},\partial_{x^{'2}}\mathfrak{D})$. Then, in view of (\ref{func est}), we obtain that, for sufficiently small $t>0$, any $(M',g')\in\mathbb{M}_t$, and any $\xi\in\mathcal{E}(M)\cap\Pi_i[\mathcal{D}]$, there exists a unique point $\xi'\in\mathcal{E}'(M')\cap\Pi_i[D]$ which is a local minimum of the function $\mathfrak{D}(\xi,\cdot)$. Moreover, the map $\alpha=\alpha_{\Pi_i[\mathcal{D}]}: \ \xi\mapsto \xi'$ is a diffeomorphism from $\mathcal{E}(M)\cap\Pi_i[\mathcal{D}]$ onto $\tilde{\alpha}(\mathcal{E}(M)\cap\Pi_i[\mathcal{D}])\subset \mathcal{E}'(M')\cap\Pi_i[D]$ satisfying 
\begin{equation}
\label{localclose 2}
\|\pi_i\circ\alpha-\pi_i\|_{C^1(\mathcal{E}(M)\cap\Pi_i[\mathcal{D}];\mathbb{C})}\le ct.
\end{equation}
In particular, (\ref{GAP est}) and (\ref{localclose 2}) imply
\begin{equation}
\label{clos p}
|\alpha(\xi)-\xi|\le ct, \qquad \forall\xi\in\mathcal{E}(M)\cap\Pi_i[\mathcal{D}].
\end{equation}
Recall that $\{U=\mathcal{E}(M)\cap\Pi_i[\mathcal{D}],\phi=\pi_i|_{U}\}$ is a holomorphic chart on $\mathcal{E}(M)$ and $\{U'=\mathcal{E}'(M')\cap\Pi_i[D],\phi'=\pi_i|_{U'}\}$ is holomorphic chart on $\mathcal{E}'(M')$. Formula (\ref{localclose 2}) yields
\begin{equation}
\label{Jac close}
\|J-I\|_{C(\mathcal{D};M_2)}\le ct,
\end{equation}
where $J=\{\partial x^{'k}/\partial x^l\}_{k,l=1,2}$ is the Jacoby matrix of the map $\alpha$. Alternatively, the map $\tilde{\alpha}=\phi'\circ\alpha\circ\phi^{-1}$ obeys $\|\partial_{z}\tilde{\alpha}-1\|_{C(\mathcal{D};\mathbb{C})}\le ct$, $\|\partial_{\overline{z}}\tilde{\alpha}\|_{C(\mathcal{D};\mathbb{C})}\le ct$. Hence, the Beltrami quotent $\mu_{\alpha}(\xi)=\partial_{\overline{z}}\tilde{\alpha}(z)/\partial_{z}\tilde{\alpha}(z)|_{z=\pi_i\xi}$ satisfies 
\begin{equation}
\label{Beltr c est}
\|\mu_{\alpha}\|_{C(U;\mathbb{C})}\le ct.
\end{equation}

$\bullet$ \ Let $\xi$ be a point of $U$ with the projection $z=\pi_i\xi$, and $\xi'=\alpha(\xi)$, $z'=\pi_i\xi'$. Let also $\theta\in T_{\xi}\mathcal{E}(M)$ be a tangent vector, and $\theta'=d\alpha_{\xi}(\theta)$. Denote by $\theta^{k}$ ($\theta^{'k}$) the components of $\theta$ ($\theta'$) in local coordinates $x^k$ ($x^{'k}$) and put $\vartheta=\theta^1+i\theta^2$ ($\vartheta'=\theta^{'1}+i\theta^{'2}$). Then $|\vartheta'\vartheta^{-1}-1|\le ct$ in view of (\ref{Jac close}). By definition of the metrics on $\mathcal{E}(M)$ and $\mathcal{E}'(M')$, we have
\begin{equation}
\label{induced m}
\begin{split}
\tilde{g}(\theta,\theta)=\rho(z)|\vartheta|^2, & \qquad \rho=\sum_{k=1}^{n}|\partial_{z}w_k\circ w_i^{-1}|^2, \\
\alpha^{*}\tilde{g}'(\theta,\theta)=\tilde{g}'(\theta',\theta')=\rho'(z')|\vartheta'|^2, & \qquad \rho'=\sum_{k=1}^{n}|\partial_{z'}w'_k\circ w_i^{'-1}|^2.
\end{split}
\end{equation}
In view of (\ref{GAP est}) and (\ref{localclose 2}), we have $\|\rho'\rho^{-1}-1\|_{C(\mathcal{D};\mathbb{C})}\le ct$. Therefore, we obtain the estimate
\begin{equation}
\label{near is}
\Big|\frac{\alpha^{*}\tilde{g}'(\theta,\theta)}{\tilde{g}(\theta,\theta)}-1\Big|\le ct, \qquad \forall \theta\in T_{\xi}\mathcal{E}(M), \ \xi\in U,
\end{equation}
which means that the map $\alpha$ is close to the isometry for small $t$.

\subsection{Construction of $\alpha$ in a zone near $\mathcal{E}(\Gamma)$}

$\bullet$ \ Now, let $\Pi_i[D]$ be an $\mathcal{E}(M)$-projective cylinder that intersects with $\mathcal{E}(\Gamma)$. Moreover, let its base $D$ be a disc with center at some point $z_0=\eta_i(l_0)$. Denote $U=\mathcal{E}(M\backslash\Gamma)\cap\Pi_i[D]$. By decreasing the radius of $D$, we obtain the following: a) $\eta_i(\Gamma)\cap D$ is a smooth curve, $D\backslash\eta_i(\Gamma)$ has two connected components $D_0$ and $D_1=\pi_i(U)$, and ${\rm mul}(w_i-z)=p$ on $D_p$; b) the function 
$$\tilde{l}: \ \eta_i^{-1}(D)\to\mathbb{R}, \qquad \tilde{l}(l):=\Re\frac{\eta_i(l)-z_0}{\partial_{\gamma}\eta_i(l_0)}$$
is invertible and the coordinates $z\mapsto\tilde{z}=(x^1,x^2)$,
\begin{equation}
\label{curverect}
x^1=\Re\frac{z-z_0}{\partial_{\gamma}\eta_i(l_0)}, \quad x^2=\Im\frac{z-\eta_i\circ\tilde{l}^{-1}(x^1)}{\partial_{\gamma}\eta_i(l_0)}
\end{equation}
are regular on $\overline{D}$. Note the `rectification property' of coordinates (\ref{curverect}): in these coordinates, the curve $\eta_i(\Gamma)\cap D$ becomes a segment of the axis $x^2=0$ (see pic. \ref{pic1}). 

Assume that $t$ is sufficiently small. Denote $U'=\mathcal{E}'(M'\backslash\Gamma)\cap\Pi_i[D]$. Applying the argument principle (\ref{GAP}) and estimate (\ref{b tr est}), we obtain that the properties a),b) are also valid for the curve $\eta'(\Gamma)$. Namely, we have a') $\eta'_i(\Gamma)\cap D$ is a smooth curve, $D\backslash\eta'_i(\Gamma)$ has two connected components $D'_0$ and $D'_1=\pi_i(U')$, and ${\rm mul}(w'_i-z)=p$ on $D'_p$; b') the function 
\begin{equation}
\label{curverect coor}
\tilde{l}': \ \eta_i^{'-1}(D)\to\mathbb{R}, \qquad \tilde{l}'(l):=\Re\frac{\eta'_i(l)-z_0}{\partial_{\gamma}\eta'_i(l_0)}
\end{equation}
is invertible and the coordinates $z'\mapsto\tilde{z}'=(x^{'1},x^{'2})$,
\begin{equation}
\label{curverect prime}
x^{'1}=\Re\frac{z'-z_0}{\partial_{\gamma}\eta'_i(l_0)}, \quad x^2=\Im\frac{z'-\eta'_i\circ\tilde{l}^{'-1}(x^1)}{\partial_{\gamma}\eta'_i(l_0)}
\end{equation}
are regular on $\overline{D}$. In particular, the cylinder $\Pi_i[D]$ is also $\mathcal{E}'(M')$-projective. Also, there exists a rectangle $\mathfrak{R}=(-a,a)\times[0,b)$, on which both functions $\tilde{z}^{-1},\tilde{z}^{'-1}$ are defined. As a corollary of (\ref{b tr est}), (\ref{curverect}), and (\ref{curverect prime}), we obtain
\begin{equation}
\label{curverect est}
\|\tilde{z}^{'-1}-\tilde{z}^{-1}\|_{C^{m}(\mathfrak{R})}+\|\tilde{l}^{'-1}-\tilde{l}^{-1}\|_{C^{m}([-a,a])}\le c_m t, \qquad m=1,2,\dots.
\end{equation}

$\bullet$ \ Now, let us prove that
\begin{equation}
\label{nearboundary1}
\|w'_j\circ w_i^{'-1}\circ\tilde{z}^{'-1}-w_j\circ w_i^{-1}\circ\tilde{z}^{-1}\|_{C^{m}(\mathfrak{R};\mathbb{C})}\stackrel{t\to 0}{\to} 0
\end{equation}
uniformly with respect to $(M',g')\in\mathbb{M}_t$. Let $\xi\in U$, $\xi'\in U'$ and their projections $z=\pi_i\xi$, $z'=\pi_i\xi'$ are connected by $\tilde{z}(z)=\tilde{z}'(z')=(x^1,x^2)\in\mathfrak{R}$. According to (\ref{GAPemb}), we have
\begin{equation}
\label{GAP near bound}
\partial^{m}_{z'}[w'_j\circ w_i^{'-1}](z')=\partial^{m}_{z'}(\pi_j\xi')=\frac{m!}{2\pi i}\int\limits_{\Gamma}\frac{\eta'_j d\eta'_i}{(\eta'_i-z')^{m+1}}.
\end{equation}
Denote $\zeta_0:=\tilde{z}^{-1}((x^1,0))$ and $\zeta'_0:=\tilde{z}^{'-1}((x^1,0))$ (see pic. \ref{pic2}). Let $\zeta'\in\eta'_i(\Gamma)$ and $\tilde{z}'(\zeta')=(y^{'1},0)\in\mathfrak{R}$. According to (\ref{curverect prime}), we have
\begin{equation}
\label{geom ineq p}
y^{'1}-x^{'1}=\Re\frac{\zeta'-\zeta'_0}{\partial_{\gamma}\eta'_i(l_0)}.
\end{equation}
From Taylor formula for $\eta'_j\circ\tilde{l}^{'-1}$ in a neighbourhood of $x^1$, we obtain
\begin{equation}
\label{Taylor ex}
|\eta'_j\circ\tilde{l}^{'-1}(y^{'1})-\mathscr{P}'_{m,x^{'1}}(\zeta')|\le c|\zeta'-\zeta'_0|^{m+1}
\end{equation}
($c$ is independent of $(M',g')$ due to (\ref{b tr est}) and (\ref{curverect est})). Here $\mathscr{P}'_{m,x^{'1}}$ is a $m$-th order polynomial in $\zeta'-\zeta'_0$ and $\overline{\zeta'-\zeta'_0}$. Comparing (\ref{curverect coor}) and (\ref{curverect prime}), we obtain $\eta'_j\circ\tilde{l}^{'-1}(y^{'1})=\eta'_j\circ\eta^{'-1}_i(\zeta')=w'_j\circ w^{'-1}_i(\zeta')$. Since $w'_j\circ w^{'-1}_i$ is holomorphic on $D'_1$, the terms with $\overline{\zeta'-\zeta'_0}$ are not included in the polynomial $\mathscr{P}'_{m,x^{'1}}$. Thus, $\mathscr{P}'_{m,x^{'1}}(\zeta')=\sum_{k=0}^{m}\mathscr{P}^{(k)}_{m,x^{'1}}(\zeta'-\zeta'_0)^{k}$ and 
\begin{equation}
\label{int ex}
\begin{split}
\partial^{m}_{z'}[w'_j\circ w_i^{'-1}](z')=&\frac{m!}{2\pi i}\Big[\int\limits_{\Gamma\backslash\Gamma_\delta}\frac{(\eta'_j-\mathscr{P}'_{m,x^{'1}}\circ\eta'_i)d\eta'_i}{(\eta'_i-z')^{m+1}}+\\
+\int\limits_{\eta_i(\Gamma_\delta)}&\frac{(\eta'_j\circ\eta^{'-1}_i-\mathscr{P}'_{m,x^{'1}})d\zeta'}{(\zeta'-z')^{m+1}}\Big]+m!\mathscr{P}^{'(m)}_{m,x^{'1}},
\end{split}
\end{equation}
where $\Gamma_\delta$ is a $\delta$-neighbourhood of the point $\eta^{-1}(\zeta_0)$. In view of (\ref{Taylor ex}),
$$\Big|\frac{\eta'_j\circ\eta^{'-1}_i-\mathscr{P}'_{m,x^{'1}}}{(\zeta'-z')^{m+1}}\Big|\le c\Big|\frac{\zeta'-\zeta'_0}{\zeta'-z'}\Big|^{m+1}.$$
Formulas (\ref{curverect prime}), (\ref{geom ineq p}), and (\ref{curverect est}) lead to the estimate 
$$|\zeta'-\zeta'_0|\le c\|\tilde{z}^{'-1}\|_{C^{m}(\mathfrak{R})} \ |y^{'1}-x^{'1}|\le c|\zeta'-z'|.$$
Thus, the second integral in (\ref{int ex}) tends to zero as $\delta\to 0$ uniformly with respect to $(M',g')\in\mathbb{M}_t$. Also, due to (\ref{b tr est}), the denominator in the first integral in (\ref{int ex}) satisfies $|(\eta'_i-z')^{m+1}|\ge c_\delta>0$ for any fixed $\delta$ and all $z'\in\tilde{z}^{'-1}(\mathfrak{R})$. Of course, the facts above are also valid if we omit the primes in (\ref{GAP near bound})-(\ref{int ex}). Also, from (\ref{curverect est}) we have 
$$|\mathscr{P}^{'(m)}_{m,x^{'1}}-\mathscr{P}^{(m)}_{m,x^{1}}|\le ct$$ 
for the coefficients in the polynomials in (\ref{Taylor ex}) in primed and non-primed cases. Due to this and (\ref{b tr est}), we have
$$\Big|\int\limits_{\Gamma\backslash\Gamma_\delta}\frac{(\eta'_j-\mathscr{P}'_{m,x^{'1}}\circ\eta'_i)d\eta'_i}{(\eta'_i-z')^{m+1}}-\int\limits_{\Gamma\backslash\Gamma_\delta}\frac{(\eta_j-\mathscr{P}_{m,x^{1}}\circ\eta_i)d\eta_i}{(\eta_i-z)^{m+1}}\Big|\le c(\delta)t$$
for any fixed $\delta$ and all $(x^1,x^2)\in\mathfrak{R}$. From these facts, formula (\ref{nearboundary1}) follows immediately.

$\bullet$ \ Recall the expression (\ref{induced m}) for the metrics $\tilde{g}$ ($\tilde{g}'$) induced by $\mathbb{C}^m$ on $\mathcal{E}(M)$ ($\mathcal{E}'(M')$) in coordinates $(\Re z,\Im z)$, $z=\pi_i\xi$ (coordinates $(\Re z',\Im z')$, $z'=\pi_i\xi'$). Due to this and formulas (\ref{curverect est}), (\ref{nearboundary1}), we obtain the estimate
\begin{equation}
\label{nearboundary2}
\|\tilde{g}'_{ij}-\tilde{g}_{ij}\|_{C^m(\mathfrak{R})}\stackrel{t\to 0}{\to} 0
\end{equation}
for the components of metric tensors $\tilde{g}$ and $\tilde{g}'$ in local coordinates $\tilde{z}$ and $\tilde{z}'$. Note that the convergence in (\ref{nearboundary2}) is uniform with respect to $(M',g')\in\mathbb{M}_t$. 

Let $(l,r)$ and $(l', r')$ stand for the semi-geodesic coordinates of points $\xi\in\mathcal{E}(M)$ and $\xi'\in\mathcal{E}'(M')$, respectively. Denote 
\begin{equation}
\label{geo neig}
\begin{split}
V&=\{\xi\in\mathcal{E}(M) \ | \ {\rm dist}_{\Gamma}(l,l_0)<q,  \ r\in[0,q)\},\\
V'&=\{\xi'\in\mathcal{E}'(M') \ | \ {\rm dist}_{\Gamma}(l',l_0)<q,  \ r'\in[0,q)\}.
\end{split}
\end{equation}
We are going to prove that there exists a sufficiently small $q>0$ independent of $(M',g')\in\mathbb{M}_t$ and such that the coordinates $(l,r)$ and $(l', r')$ are regular on $V$ and $V'$, respectively. Also, we will prove that the map $\alpha_V: \ V\to V'$, defined by the rule
$$l'(\alpha_V(\xi))=l(\xi), \quad r'(\alpha_V(\xi))=r(\xi),$$
is near-isometric for small $t$. To this end, we need the following auxiliary fact.

Let $\mathscr{B}$ be a neighbourhood of the origin in the half-plane $\mathbb{R}\times[0,+\infty)$ containing a segment $[-r_0,r_0]\times\{0\}$ and $h$ be a metric tensor on $\mathscr{B}$ with components $h_{ij}\in C^2(\mathscr{B};\mathbb{R})$. Denote the outward normal on $[-r_0,r_0]\times\{0\}$ corresponding to the metric $h$ by $\nu_h$. Introduce the bundle $x_h$ of semi-geodesics (with respect to the metric $h$) curves $r\mapsto x_h(r,\mu)$ in $\mathscr{B}$, where $\mu\in [-r_0,r_0]$, $r>0$, and
$$x_h(0,\mu)=(0,\mu), \quad \partial_r x_h(0,\mu)=-\nu_h(\mu).$$
Put $\mathscr{Q}_{r}=(-r,r)\times[0,r)$.
\begin{lemma}
\label{geo-lemma}
Suppose that $x_g$ is a diffeomorphism from $\mathscr{Q}_{r_0}$ onto $x_h(\mathscr{Q}_{r_0})\subset\mathscr{B}$. Then there exist $s_0>0$ and the sub-rectangle $\mathscr{Q}_{r_1}\subset \mathscr{Q}_{r_0}$ such that, for $s\in (0,s_0)$ and any metrics $h'$ on $\mathscr{B}$, satisfying
\begin{equation}
\label{mt-clos}
\|h'_{ij}-h_{ij}\|_{C^2(\mathscr{B};\mathbb{R})}\le s,
\end{equation}
the map $x_{h'}$ is a diffeomorphism from $\mathscr{Q}_{r_1}$ onto $x_{h'}(\mathscr{Q}_{r_1})\subset\mathscr{B}$ and
\begin{equation}
\label{geo-clos}
\|x_{h'}-x_{h}\|_{C^1(\mathscr{Q}_{r_1};\mathbb{R}^2)}\le cs.
\end{equation}
In particular, the map $\varkappa:=x_{h'}\circ x_{h}^{-1}$ obeys
\begin{equation}
\label{geo near is}
\|(\varkappa^{*}h')_{ij}-h_{ij}\|_{C(\mathscr{Q}_{r_1};\mathbb{R})}\le cs.
\end{equation}
\end{lemma}
Lemma \ref{geo-lemma} is just a variant of the theorem on the local solvability of Cauchy problem; for the convenience of the reader, this proof is contained in Appendix \ref{App2}.

Put $\mathscr{B}=\mathfrak{R}$ and $h_{ij}=\tilde{g}_{ij}$, $h'_{ij}=\tilde{g}'_{ij}$. According to Lemma \ref{geo-lemma} and formula (\ref{nearboundary2}), the semi-geodesics coordinates $(l,r)$ and $(l', r')$ are regular on the neighbourhoods $V$ and $V'$ given by (\ref{geo neig}) for sufficiently small $t>0$ and $q=q(l_0)>0$. For the map $\alpha_V$, formula (\ref{geo near is}) takes the form
\begin{equation}
\label{nearboundary mt}
\|(\alpha_V^{*}\tilde{g}')_{ij}-\tilde{g}_{ij}\|_{C(\tilde{z}\circ\pi_i(V))}\stackrel{t\to 0}{\to} 0,
\end{equation}
where $\tilde{g}_{ij}$ and $(\alpha_V^{*}\tilde{g}')_{ij}$ are the components of metric tensors $\tilde{g}$ and $\alpha_V^{*}\tilde{g}'$ in local coordinates $\tilde{z}$.
In view of (\ref{curverect est}), formula (\ref{geo-clos}) yields
\begin{equation}
\label{nearboundary map}
\|\pi_i\circ\alpha_V-\pi_i\|_{C^{1}(V;\mathbb{C})}\stackrel{t\to 0}{\to} 0.
\end{equation}
Chose the holomorphic charts $(\pi_i(V),\phi=\pi_i|_V)$ and $(\pi_i(V'),\phi'=\pi_i|_V')$ on $\mathcal{E}(M)$ and $\mathcal{E}'(M')$, respectively. Due to (\ref{nearboundary map}), the map $\tilde{\alpha}_V=\phi'\circ\alpha_V\circ\phi^{-1}$ obeys $\|\partial_{z}\tilde{\alpha}_V-1\|_{C(\pi_i(V);\mathbb{C})}\le ct$, $\|\partial_{\overline{z}}\tilde{\alpha}_V\|_{C(\pi_i(V);\mathbb{C})}\le ct$. Hence, its Beltrami quotent $\mu_{\alpha_V}(\xi)=\partial_{\overline{z}}\tilde{\alpha}_V(z)/\partial_{z}\tilde{\alpha}_V(z)|_{z=\pi_i\xi}$ satisfies 
\begin{equation}
\label{Beltr c est nera b}
\|\mu_{\alpha_V}\|_{C(\pi_i(V);\mathbb{C})}\stackrel{t\to 0}{\to} 0.
\end{equation}
Estimates (\ref{nearboundary mt}), (\ref{nearboundary map}), and (\ref{Beltr c est nera b}) are uniform with respect to $(M',g')\in\mathbb{M}_t$.

The neighbourhoods (\ref{geo neig}) with $l_0\in\Gamma$ and $q=q(l_0)$ provide an open cover of the compact $\mathcal{E}(\Gamma)$ in $\mathcal{E}(M)$. Chose a finite sub-cover $\{V_k\}_{k=1}^{L}$ and the positive $r_0>0$ such that the near-boundary strip $r(\xi)\in[0,r_0)$ is contained in the union of $V_k$. Chose the cut-off function $\kappa$ in (\ref{d whole}) in such a way that $\kappa=0$ for $r>2r_0 /3$ and $\kappa=1$ for $r<r_0 /3$. By definition, the map $\alpha$ satisfies
$$l'(\alpha(\xi))=l(\xi), \quad r'(\alpha(\xi))=r(\xi)$$
for any $\xi$ in the strip $r(\xi)<r_0 /3$ and, hence, $\alpha(\xi)=\alpha_V(\xi)$ for each $V\ni\xi$. Then estimate (\ref{nearboundary mt}) yields 
\begin{equation}
\label{near is near b}
\sup_{\theta\in T_{\xi}\mathcal{E}(M), \ r(\xi)\in[0,r_0)}\Big|\frac{\alpha^{*}\tilde{g}'(\theta,\theta)}{\tilde{g}(\theta,\theta)}-1\Big|\stackrel{t\to 0}{\to} 0.
\end{equation}
Estimate (\ref{nearboundary map}) implies
\begin{equation}
\label{p clos nb}
\sup_{r(\xi)\in[0,r_0)}|\alpha(\xi)-\xi|\stackrel{t\to 0}{\to} 0.
\end{equation}
Note that both convergences (\ref{near is near b}) and (\ref{p clos nb}) are uniform with respect to $(M',g')\in\mathbb{M}_t$.

\subsection{Construction of $\alpha$ in an intermediate zone}
\label{ss inter}

$\bullet$ \ Let $\Pi_i[D]$ be an $\mathcal{E}(M)$-projective cylinder whose intersection $\mathcal{E}(M)\cap\Pi_i[D]$ is contained in the strip $r\in(r_0/6,5r_0/6)$. For $\xi\in \mathcal{E}(M)\cap\Pi_i[D]$, we have $\kappa(\xi)\in (0,1)$ and the functions $\mathfrak{D}$ and $\mathfrak{D}_0$ in (\ref{d whole}) and (\ref{d unpert}) are of general type. Nonetheless, due to estimates (\ref{nearboundary map}) and (\ref{nearboundary1}), we still have the (uniform with respect to $(M',g')\in\mathbb{M}_t$) estimate
$$\|\mathfrak{D}-\mathfrak{D}_0\|_{C^1(\mathcal{E}(M)\cap\Pi_i[D];[0,+\infty))}\stackrel{t\to 0}{\to} 0.$$
Also, recall that the function $\xi'\mapsto\mathfrak{D}_0(\xi,\xi')$ has a unique and non-degenerated global minimum point $\xi'=\xi$. So, by repeating of arguments of Subsection \ref{ss out of b} (including the application of Lemma \ref{IFLemma}), we obtain that the map $\alpha$ is well defined on $U=\mathcal{E}(M)\cap\Pi_i[\mathcal{D}]$, where $\Pi_i[\mathcal{D}]$ is a sub-cylinder of $\Pi_i[D]$. Also, 
\begin{equation}
\label{p clos}
\sup_{\xi\in U}|\alpha(\xi)-\xi|\stackrel{t\to 0}{\to} 0,
\end{equation}
\begin{equation}
\label{near is inter}
\sup_{\theta\in T_{\xi}\mathcal{E}(M), \ \xi\in U}\Big|\frac{\alpha^{*}\tilde{g}'(\theta,\theta)}{\tilde{g}(\theta,\theta)}-1\Big|\stackrel{t\to 0}{\to} 0
\end{equation}
and the map $\tilde{\alpha}=\phi'\circ\alpha\circ\phi^{-1}$ (where $\phi=\pi_i|_{\mathcal{E}(M)\cap\Pi_i[D]}$ and $\phi'=\pi_i|_{\mathcal{E}'(M')\cap\Pi_i[D]}$) satisfies 
\begin{equation}
\label{Beltr c est inter}
\|\mu_{\alpha}\|_{C(\pi_i(U);\mathbb{C})}\stackrel{t\to 0}{\to} 0,
\end{equation}
where $\mu_{\alpha}(\xi)=\partial_{\overline{z}}\tilde{\alpha}(z)/\partial_{z}\tilde{\alpha}(z)|_{z=\pi_i\xi}$ is the Beltrami quotent of $\tilde{\alpha}$. Note that all convergences (\ref{p clos})-(\ref{Beltr c est inter}) are uniform with respect to $(M',g')\in\mathbb{M}_t$.

\subsection{Construction of the global diffeomorphism $\alpha$}
$\bullet$ \ Consider the open cover of the compact $\mathcal{E}(M)$ by the following sets: a) the near-boundary strip $U_{0}=\{\xi\in\mathcal{E}(M) \ | \ r(\xi)\in[0,r_0/3)\}$, b) domains $U=\mathcal{E}(M)\cap\Pi_i[\mathcal{D}]$, where $\Pi_i[\mathcal{D}]$ is a sub-cylinder of $\mathcal{E}(M)$-projective cylinder $\Pi_i[D]$ which does not intersect the strip $r\in[0,r_0/6]$ in $\mathcal{E}(M)$. Fix some its finite sub-cover $\{U_0,U_1,\dots,U_K\}$ and denote the cylinders corresponding to $U_k$, $k=1,\dots,K$ by $\Pi_{i_k}[\mathcal{D}_k]$ and $\Pi_{i_k}[D_k]$. According to the Subsections \ref{ss out of b}-\ref{ss inter}, the map $\alpha=\alpha_k$ is defined locally on each $U_k$.

Suppose that $\xi\in U_{k}\cap U_{s}$, where $k\ne 0$ and $s\ne k$. Put $\xi'=\alpha_k(\xi)$, $\xi''=\alpha_s(\xi)$. Due to estimates (\ref{clos p}), (\ref{p clos nb}), (\ref{p clos}), for any $\delta>0$, there exists $t(\delta)>0$ such that $|\xi''-\xi'|<\delta$ for any $t\in(0,t(\delta))$. Hence, for sufficiently small $t$, the point $\xi''$ belongs to $\mathcal{E}'(M')\Pi_{i_k}[D_k]$. Thus, $\xi'$ and $\xi''$ are both points of minimum of the function $\mathfrak{D}(\xi,\cdot)$ in $\mathcal{E}'(M')\Pi_{i_k}[D_k]$. Due to results of Subsections \ref{ss out of b} and \ref{ss inter}, the function $\mathfrak{D}(\xi,\cdot)$ has a unique point of minimum in $\mathcal{E}'(M')\Pi_{i_k}[D_k]$. Thus, $\xi''=\xi'$. So, the map $\alpha$ is well-defined globally on $\mathcal{E}(M)$ for sufficiently small $t$. By, definition, $\alpha\circ\mathcal{E}(l)=\mathcal{E}(l')$ for $l\in\Gamma$.

Now, suppose that $\alpha(\xi_1)=\alpha(\xi_2)=\xi'$, where $\xi_1\ne\xi_2$. Since $\alpha|_{U_k}$ is a diffeomorphism of $U_k$ and $\alpha(U_k)$ for each $k$, there is no domains $U_k$ containing both $\xi_1$ and $\xi_2$. Hence, $|\xi_2-\xi_1|\ge c>0$. Otherwise, estimates (\ref{clos p}), (\ref{p clos nb}), (\ref{p clos}) imply that $|\xi_2-\xi_1|<c/2$ for sufficiently small $t$, a contradiction. Thus, the map $\alpha$ is an injection and, thus, a diffeomorphism from $\mathcal{E}(M)$ to $\alpha(\mathcal{E}(M))$. In particular, the set $\alpha(\mathcal{E}(M))$ is open in $\mathcal{E}(M)$. Let $\alpha(\xi_s)\to \xi'$ as $s\to\infty$. Since $\mathcal{E}(M)$ is compact, there is a subsequence of $\{\xi_s\}$ which converges to some point $\xi$ in $\mathcal{E}(M)$ (for simplicity, let this subsequence coincides with $\{\xi_s\}$). Since $\alpha$ is continuous, we have $\alpha(\xi_s)\to \alpha(\xi)$ as $s\to\infty$ and, hence, $\xi'=\alpha(\xi)$. Thus, the set $\alpha(\mathcal{E}(M))$ is also closed in $\mathcal{E}(M)$. So, we have $\alpha(\mathcal{E}(M))=\mathcal{E}'(M')$ and the map $\alpha=\alpha_{M'}$ is a diffeomorphism from $\mathcal{E}(M)$ onto $\mathcal{E}'(M')$ for small $t$ and all $(M',g')\in\mathbb{M}_t$.

Estimates (\ref{near is}),(\ref{nearboundary mt}),(\ref{near is inter}) imply that
$$\sup_{\theta\in T_{\xi}\mathcal{E}(M), \ \xi\in \mathcal{E}(M)}\Big|\frac{\alpha^{*}\tilde{g}'(\theta,\theta)}{\tilde{g}(\theta,\theta)}-1\Big|\stackrel{t\to 0}{\to} 0$$
uniformly with respect to $(M',g')\in\mathbb{M}_t$. So, the map $\alpha$ is near isometric for small $t$. 

Introduce the diffeomorphism $\sigma_{M'}=\mathcal{E}^{'-1}\circ\alpha_{M'}\circ\mathcal{E}$. In view of (\ref{Beltr c est}), (\ref{Beltr c est nera b}), (\ref{Beltr c est inter}), the map $\sigma_{M'}: \ (M,g)\to(M',g')$ is $K$-quasi-conformal with the dilatation $K=K(\sigma_{M'})$ satisfying 
$$\sup_{(M',g')\in\mathbb{M}_t}K(\sigma_{M'})\stackrel{t\to 0}{\to} 1.$$
In view of the definition (\ref{TM}) of Teichm\"uller distance between $[(M,g)]$ and $[(M',g')]$, we obtain
$$\sup_{(M',g')\in\mathbb{M}_t}d_T([(M,g)],[(M',g')])\le\frac{1}{2}{\rm log}\sup_{(M',g')\in\mathbb{M}_t}\inf_{q}K(\sigma_{M'})\stackrel{t\to 0}{\to} 0.$$
So, we have proved (\ref{MTf}) and Theorem \ref{MT}.

\appendix

\section{Proof of Lemma \ref{IFLemma}}
\label{App1}
Due to condition ii, the function $x\to |{\rm det}F'_y(x,f(x)|$ attains the positive minimum $m_0$ on the compact $\overline{\mathcal{X}}$. Hence, in view of (\ref{FG C1-closeness}), we have $|{\rm det}F'_y(x,f(x)|>m_0/2$ for sufficiently small $t$ and any $x\in\overline{\mathcal{X}}$. Introduce the functions
\begin{align*}
%\label{contraction F}
\mathcal{F}(x,y):=y-(F'_y(x,f(x))^{-1}F(x,y),\\
%\label{contraction G}
\mathcal{H}(x,y):=y-(H'_y(x,f(x))^{-1}H(x,y);
\end{align*}
then $(x,y)$ is a zero of $F$ (or $H$) if and only if $y$ is a fixed point of the map $\mathcal{F}(x,\cdot)$ (or $\mathcal{H}(x,\cdot)$). Note that
\begin{align}
\label{contraction F derivative}
\mathcal{F}'_{y}(x,y)=I-(F'_y(x,f(x))^{-1}F'_y(x,y),\\
\label{contraction G derivative}
\mathcal{H}'_{y}(x,y)=I-(H'_y(x,f(x))^{-1}H'_y(x,y)
\end{align}
and, in view of (\ref{FG C1-closeness}),
\begin{equation}
\| \mathcal{H}'_{y}(x,\cdot)-\mathcal{F}'_{y}(x,\cdot)\|_{C(\mathcal{X};Y)}\le ct
\end{equation}
(here and in the subsequent, the constants do not depend on $H$).

\smallskip

For $x\in X$, $y\in Y$, and $\varepsilon>0$, denote the $\varepsilon$-neighbourhood of $x$ ($y$) in $X$ ($Y$) by $U_\varepsilon(x)$ ($V_\varepsilon(y)$). Let $x_0\in \mathcal{X}$ and $y_0=f(x_0)$, then $F(x_0,y_0)=0$ and $\mathcal{F}'_{y}(x_0,y_0)=0$. Since $F\in C^1(X,Y)$, for any $\delta>0$, there exists a sufficiently small $\varepsilon=\varepsilon(\delta)>0$ such that $|\mathcal{F}'_{y}(x,y)|\le \delta$ for any $x\in U_\varepsilon(x_0)$ and $y\in V_\varepsilon(f(x))$. In view of (\ref{FG C1-closeness}), there exists $t_0=t_0(\delta)>0$ such that $|\mathcal{H}'_{y}(x,y)|\le 2\delta$ for any $t\in (0,t_0)$ and any $x\in U_\varepsilon(x_0)$ and $y\in V_\varepsilon(f(x))$. Then, choosing sufficiently small $\delta=\delta(x_0)$, we obtain
\begin{align*}
|\mathcal{H}&(x,y')-\mathcal{H}(x,y)|\le c\delta|y'-y|<|y'-y|,\\
|\mathcal{H}&(x,y)-\mathcal{H}(x,f(x))|\le c\delta\varepsilon<\varepsilon
\end{align*}
for $t\in (0,t_0)$, $x\in U_\varepsilon(x_0)$, and $y\in V_\varepsilon(f(x))$. So, the map $\mathcal{H}(x,\cdot): \ V_\varepsilon(f(x))\mapsto V_\varepsilon(f(x))$ is a contraction. Due to the Banach fixed-point theorem, for each $x\in U_\varepsilon(x_0)$ and $t\in (0,t_0)$, there exists a unique point $y=:h(x)\in V_\varepsilon(f(x))$ such that $\mathcal{H}(x,y)=h(x)$ i.e. $H(x,y)=0$. Also, the condition $\|\mathcal{H}'_{y}(x,y)\|\le 2\delta$ and formula (\ref{contraction G derivative}) imply that $H'_y(x,y)^{-1}$ exists for any $t\in (0,t_0)$, $x\in U_\varepsilon(x_0)$, and $y\in V_\varepsilon(f(x))$; in particular, there exists $H'_y(x,g(x))^{-1}$. Therefore, since $H\in C^1(X\times Y;\mathbb{R}^n)$, we obtain $h\in C^1(U_\varepsilon(x_0);Y)$ from the implicit function theorem.

\smallskip

In view of (\ref{FG C1-closeness}) and the following equality
\begin{align*}
0=H(x,h(x))-F(x,f(x))&=\\
=H(x,h(x))-F(x,&h(x))+F(x,h(x))-F(x,f(x)),
\end{align*}
we have 
$$|\int_{0}^{\tau}F'_y(x,f(x)+es)eds|=|F(x,h(x))-F(x,f(x))|\le ct,$$
where $\tau=|h(x)-f(x)|$ and $e=(h(x)-f(x))/\tau$. Also, due to the condition $\|\mathcal{F}'_{y}(x,y)\|\le \delta$ and formula (\ref{contraction F derivative}), we have $\|F'_y(x,y)-F'_y(x,f(x))\|\le c\delta$ for $t\in (0,t_0)$, $x\in U_\varepsilon(x_0)$, and $y\in V_\varepsilon(f(x))$. Thus,
$$|\int_{0}^{\tau}F_y(x,f(x)+es)eds|\ge \tau |F_y(x,f(x))e|-c\delta\tau\ge c\tau,$$
whence $\tau\le ct$. So, we have proved that $\|h-f\|_{C(U_\varepsilon(x_0));Y)}\le ct$. Next, 
\begin{align*}
0&=\frac{dF(x,f(x))}{dx}=F'_x(x,f(x))+F'_y(x,f(x))f'(x),\\
0&=\frac{dH(x,h(x))}{dx}=H'_x(x,h(x))+H'_y(x,h(x))h'(x),
\end{align*}
whence
\begin{align*}
f'(x)-h'(x)=(H'_y(x,h(x)))^{-1}H'_x(x,h(x))-(F'_y(x,f(x)))^{-1}F'_x(x,f(x))=\\=(H'_y(x,h(x)))^{-1}H'_x(x,h(x))-(F'_y(x,h(x)))^{-1}F'_x(x,h(x))+\\+(F'_y(x,h(x)))^{-1}F'_x(x,h(x))-(F'_y(x,f(x)))^{-1}F'_x(x,f(x)).
\end{align*}
Thus, $\|h-f\|_{C^1(U_\varepsilon(x_0));Y)}\le ct$ since $F\in C^2(\overline{X}\times \overline{Y};\mathbb{R}^n)$ and the condition (\ref{FG C1-closeness}) holds.

\smallskip

The neighbourhoods $U_{\varepsilon(\delta)}(x_0)$ ($x_0 \in \mathcal{X}$, $\delta=\delta(x_0)$) provides an open cover of $\mathcal{X}$. Choose a finite subcover $U_{\varepsilon(\delta)}(x_{0,k})$ ($k=1,\dots,K$) and put $t_*=\min\{t_0(\delta(x_{0,k}))\}_k$. Then, for $t\in (0,t_*)$, the function $h$ is defined globally on $\mathcal{X}$ and obeys $H(x,h(x))=0$ for $x\in \mathcal{X}$ and satisfies (\ref{FG C1-closeness}). Also, there exists neighbourhood $N$ of the set $\{(x,f(x)) \ | \ x\in \mathcal{X}\}$ in $\overline{\mathcal{X}}\times \overline{Y}$ such that all zeroes of $H$ in $N$ belong to the graph of $h$. The set $N'=(\overline{\mathcal{X}}\times \overline{Y})\backslash N$ is compact, so $|F|$ attains positive minimum $M$ on $N'$. Due to (\ref{FG C1-closeness}), $H$ has no zeroes on $N'$ for sufficiently small $t>0$.

\smallskip

Now, suppose that $f$ is a diffeomorphism from $X$ to $f(X)$. Therefore, $|{\rm det}f'(x)|>c_0$ and $|f(x')-f(x)|\ge c_1{\rm dist}_{\overline{\mathcal{X}}}(x',x)$ ($c_0,c_1>0$) for any $x\in \overline{\mathcal{X}}$. Due to (\ref{FG C1-closeness}), $|{\rm det}h'(x)|>c_0/2>0$ for $x\in \mathcal{X}$. Suppose that $h(x)=h(x')$, then
\begin{align*}
0=\int_{x}^{x'}h'(s)ds=\int_{x}^{x'}f'(s)ds+&\int_{x}^{x'}(h'(s)-f'(s))ds=\\
=&f(x')-f(x)+\int_{x}^{x'}(h'(s)-f'(s))ds,
\end{align*}
where integral is taken over shortest curve in $\overline{\mathcal{X}}$ connecting $x$ and $x'$. Thus,
$$c_1{\rm dist}_{\overline{\mathcal{X}}}(x',x)\le |\int_{x}^{x'}(h'(s)-f'(s))ds|\le ct{\rm dist}_{\overline{\mathcal{X}}}(x',x)$$
due to (\ref{FG C1-closeness}). So, for sufficiently small $t$, the equality $h(x)=g(x')$ implies $x'=x$. Therefore $h$ is a a diffeomorphism from $\mathcal{X}$ to $h(\mathcal{X})$.

\section{Proof of Lemma \ref{geo-lemma}}
\label{App2}

Denote $X_1:=x_h$, $X_2:=\partial_r x_h$ and $Y_1:=x_{h'}$, $Y_2:=\partial_r x_{h'}$. The functions $X=(X_1,X_2)$ and $Y=(Y_1,Y_2)$ satisfy the following Cauchy problems for geodesic equations
\begin{align}
\label{geo CP 1}
\frac{\partial X}{\partial r}&=A(X), \quad X(0,\mu)=X_0(\mu), \\
\label{geo CP 2}
\frac{\partial Y}{\partial r}&=B(Y), \quad Y(0,\mu)=Y_0(\mu),
\end{align}
where 
\begin{align*}
A(X)&=(X^2,-\Gamma^{i}_{jk}(X^1)X_2^j X_2^k), & X_0(\mu)=((0,\mu),-\nu_h(\mu)) \\
B(Y)&=(Y^2,-\tilde{\Gamma}^{i}_{jk}(Y^1)Y_2^j Y_2^k), & Y_0(\mu)=((0,\mu),-\nu_{h'}(\mu))
\end{align*}
and $\Gamma^{i}_{jk}$ and $\tilde{\Gamma}^{i}_{jk}$ are Christoffel symbols corresponding to metrics $h$ and $h'$, respectively.
Denote by $\mathscr{V}$ a bounded domain in $\mathbb{R}^2$ containing all $-\nu_h(\mu)$ ($\mu\in(-r_0,r_0)$). Condition (\ref{mt-clos}) implies
\begin{equation}
\label{AB ests}
\|A-B\|_{C^1(\mathscr{B}\times\mathscr{V})}\le cs, \quad \|Y_0-X_0\|_{C^1([-r_0,r_0])}\le cs.
\end{equation}
The solutions of (\ref{geo CP 1}) and (\ref{geo CP 2}) are fixed points of the maps
\begin{align*}
\mathcal{A}[X](r,\mu)=X_0(\mu)+\int\limits_0^r A(X(r',\mu))dr',\\
\mathcal{B}[Y](r,\mu)=Y_0(\mu)+\int\limits_0^r B(Y(r',\mu))dr'.
\end{align*}
Put $a:=\|A\|_{C^1(\mathscr{B}\times\mathscr{V})}$. For $r_1>0$ and $\epsilon>0$, denote the closed ball in $C(\overline{\mathscr{Q}_{r_1}})$ of radius $\epsilon$ with center at $F$  by $\mathfrak{B}_{\epsilon,r_1}(F)$. According to (\ref{AB ests}), the inequalities
\begin{align*}
\|B(Y_1)-A(Y_1)\|_{C(\overline{\mathscr{Q}_{r_1}})}&\le c(1+\epsilon)s, \\
\|\mathcal{A}[X_1]-X_0\|_{C(\overline{\mathscr{Q}_{r_1}})}&\le cr_1\epsilon,\\
\|\mathcal{B}[Y_1]-Y_0\|_{C(\overline{\mathscr{Q}_{r_1}})}&\le c(a+s)r_1\epsilon
\end{align*}
and 
\begin{align*}
\|\mathcal{A}[X_1]-\mathcal{A}[X_2]\|_{C(\overline{\mathscr{Q}_{r_1}})}\le &c\|X_1-X_2\|_{C(\overline{\mathscr{Q}_{r_1}})}ar_1,\\ 
\|\mathcal{B}[Y_1]-\mathcal{B}[Y_2]\|_{C(\overline{\mathscr{Q}_{r_1}})}\le &c\|Y_1-Y_2\|_{C(\overline{\mathscr{Q}_{r_1}})}(a+s){r_1}
\end{align*}
hold for any $X_1,X_2\in\mathfrak{B}_{\epsilon,r_1}(X_0)$ and $Y_1,Y_2\in\mathfrak{B}_{\epsilon,r_1}(Y_0)$. Thus, for sufficiently small $\epsilon,r_1,s$, the maps $\mathcal{A}$ ($\mathcal{B}$) are contractions on $\mathfrak{B}_{\epsilon,r_1}(X_0)$ and $\mathfrak{B}_{\epsilon/2,r_1}(Y_0)\subset\mathfrak{B}_{\epsilon,r_1}(X_0)$, respectively, and their fixed points $X$, $Y$ satisfy
\begin{align*}
\|Y-X\|_{C(\overline{\mathscr{Q}_{r_1}})}=&\|\mathcal{B}[Y]-\mathcal{A}[X]\|_{C(\overline{\mathscr{Q}_{r_1}})}\le\|\mathcal{B}[Y]-\mathcal{A}[Y]\|_{C(\overline{\mathscr{Q}_{r_1}})}+\\
+&\|\mathcal{A}[Y]-\mathcal{A}[X]\|_{C(\overline{\mathscr{Q}_{r_1}})}\le cs+\|Y-X\|_{C(\overline{\mathscr{Q}_{r_1}})}/2.
\end{align*}
Hence $\|Y-X\|_{C(\overline{\mathscr{Q}_{r_1}})}\le cs$ and 
\begin{align*}
\|\partial_r Y-\partial_r X\|_{C(\overline{\mathscr{Q}_{r_1}})}=&\|B[Y]-A[X]\|_{C(\overline{\mathscr{Q}_{r_1}})}\le \\
\le \|B[Y]-A[Y]\|_{C(\overline{\mathscr{Q}_{r_1}})}&+\|A[Y]-A[X]\|_{C(\overline{\mathscr{Q}_{r_1}})}\le \\
&\le c(s+\|Y-X\|_{C(\overline{\mathscr{Q}_{r_1}})})\le cs.
\end{align*}
Differentiation of (\ref{geo CP 1}) and (\ref{geo CP 2}) yields the following systems of linear equations
\begin{align*}
\frac{\partial}{\partial r}\Big(\frac{\partial X}{\partial \mu}\Big)&=A'_X(X)\frac{\partial X}{\partial \mu}, \quad \frac{\partial X}{\partial \mu}(0,\mu)=(X_0)'_\mu(\mu), \\
\frac{\partial}{\partial r}\Big(\frac{\partial Y}{\partial \mu}\Big)&=B'_Y(Y)\frac{\partial Y}{\partial \mu}, \quad \frac{\partial Y}{\partial \mu}(0,\mu)=(Y_0)'_\mu(\mu).
\end{align*}
Thus, the estimates above lead to
$$\|\frac{\partial Y}{\partial \mu}-\frac{\partial X}{\partial \mu}\|_{C(\overline{\mathscr{Q}_{r_1}})}\le cs.$$
and, hence, to (\ref{geo-clos}). The remaining statements of Lemma \ref{geo-lemma} easily follows from (\ref{geo-clos}).

\subsection*{Statements and Declarations}
\paragraph{Competing Interests.} On behalf of all authors, the corresponding author states that there is no conflict of interest.

\paragraph{Data Availibility Statement.} Data sharing not applicable to this article as no datasets were generated oranalysed during the current study.

\paragraph{Funding.} M. I. Belishev and D. V. Korikov were supported by the RFBR grant
20-01-00627-a.

\end{document}